\documentclass[a4paper,11pt]{article}
\usepackage{a4wide,amsmath,amssymb,slashed,bbm,bm}
\usepackage[english]{babel}

\usepackage{cite}

\makeatletter
\g@addto@macro\bfseries{\boldmath}

\@addtoreset{equation}{section}
\makeatother

\begin{document}

\hfill{ 
}

\vspace{30pt}

\begin{center}
{\huge{\bf Classifying integrable symmetric space strings via factorized scattering}}

\vspace{50pt}

{\bf Linus Wulff}

\vspace{15pt}

{\it\small Department of Theoretical Physics and Astrophysics, Masaryk University, 611 37 Brno, Czech Republic}\\

\vspace{100pt}

{\bf Abstract}
\end{center}
\noindent
All symmetric space $AdS_n$ solutions of type II supergravity have recently been found for $n>2$. For the supersymmetric solutions (and their T-duals) it is known that the Green-Schwarz string is classically integrable. We complete the classification by ruling out integrability for the remaining non-supersymmetric solutions. This is achieved by showing that tree-level scattering on the worldsheet of a GKP or BMN string fails to factorize for these cases.

\pagebreak 
\tableofcontents

\setcounter{page}{1}


\section{Introduction and main results}
Two-dimensional symmetric space sigma models constitute a classic example of integrable models \cite{Pohlmeyer:1975nb,Eichenherr:1979ci}. 
Since in string theory a critical string is also described by a 2d sigma model one might think that if the string is moving in a background which happens to be a symmetric space the string sigma model should be classically integrable. This expectation is to naive however for two reasons: 
\begin{itemize}
	\item[(i)] besides the standard symmetric space sigma model action the string action contains also a coupling to the two-form B-field and it is not clear in general that this coupling is compatible with the integrability,
	\item[(ii)] the string action contains fermions as well as bosons and it is not clear in general that the integrability of the bosonic sector extends to the fermions.
\end{itemize}
Regarding (ii) it is sometimes claimed that in the Green-Schwarz formulation of the string, which is the one we will be using here, the fermionic kappa symmetry transforms bosons and fermions into each other and therefore integrability for the bosons is enough to guarantee integrability for the full theory. This is wrong for the following reason. Despite the name kappa symmetry is just a gauge invariance of the Green-Schwarz formulation whose only role is to remove half of the would be fermionic degrees of freedom -- it does \emph{not} relate the \emph{physical} bosonic and fermionic degrees of freedom to each other. However, if there is supersymmetry present then bosonic and fermionic degrees of freedom are related and it is natural to expect integrability of the bosonic sector to carry over to the fermions. Indeed it was shown in \cite{Wulff:2015mwa} that for a supersymmetric symmetric space RR background (NSNS flux would run into issue (i) above) the full superstring is classically integrable by directly constructing the Lax connection to quadratic order in fermions. Conversely, there seems to be no good reason to expect the integrability to be there in the absence of supersymmetry.\footnote{This has a somewhat similar flavor to arguments that non-supersymmetric AdS solutions should be unstable \cite{Ooguri:2016pdq,Freivogel:2016qwc}. Our findings here are consistent with this conjecture in the sense that we do not find any genuinely non-supersymmetric solutions which are integrable (while such solutions would not directly contradict the instability conjecture their existence would be very puzzling if the conjecture were true).}

Here we will attempt a classification of the (classically) integrable symmetric space strings. Since, at the classical level, the target space geometry of the string must solve the supergravity equations of motion we need to know all symmetric space solutions of the supergravity equations. We will focus here on type II supergravity which is the most interesting case. It is not hard to show that there are only two types of symmetric space solution possible \cite{Figueroa-OFarrill:2012whx}: pp-waves (or Cahen-Wallach spaces) or $AdS_n\times M_{10-n}$ with $M_{10-n}$ a Riemannian symmetric space. For the first class the Green-Schwarz string action becomes quadratic in light-cone gauge and hence trivially integrable. These backgrounds also always preserve at least half the supersymmetries (i.e. 16) consistent with our argument above that supersymmetry should be a necessary condition for integrability.

We are therefore left with the problem of analyzing the integrability of strings in symmetric $AdS_n\times M_{10-n}$ backgrounds. Such backgrounds were classified, for $n>2$, in \cite{Wulff:2017zbl} following \cite{Figueroa-OFarrill:2012whx} which analyzed the type IIB case. Out of these some are known to be integrable, namely all the supersymmetric solutions plus two non-supersymmetric ones. These are indicated with a 'Y' in the last column in table \ref{tab:backgrounds}.\footnote{A classification of so-called $\mathbbm Z_4$-symmetric supercoset sigma models with vanishing beta-function was given in \cite{Zarembo:2010sg}. Many of these arise as truncations of the models corresponding to the solutions in the table.} The classical integrability of the string in $AdS_5\times S^5$ was established in \cite{Bena:2003wd} and the integrability of the $AdS_5\times\mathbbm{CP}^2\times S^1$ string follows from the observation that it is T-dual to the former \cite{Duff:1998us}. Note that the T-duality on the Hopf fiber of $S^5$ breaks the supersymmetry at the supergravity level, however the string still knows about this symmetry which is now realized non-locally on the worldsheet. When we say that supersymmetry seems to be needed for integrability it does not have to be present at the supergravity level but can be hidden as in this case. Integrability of the $AdS_4\times\mathbbm{CP}^3$ string was established by the supercoset construction of the action in \cite{Arutyunov:2008if,Stefanski:2008ik} and then for the complete Green-Schwarz string \cite{Gomis:2008jt} in \cite{Sorokin:2010wn} by constructing a Lax connection to quadratic order in all fermions from components of conserved isometry currents and exactly for a certain truncated model (see also \cite{Cagnazzo:2011at}). For $AdS_3\times S^3\times S^3\times S^1$ and $AdS_3\times S^3\times T^4$, which can be obtained as a limit of the former, the integrable supercoset model was described in \cite{Babichenko:2009dk} and in the case of mixed NSNS and RR flux in \cite{Cagnazzo:2012se}, while the full superstring construction to quadratic order in fermions was done in \cite{Sundin:2012gc} and \cite{Wulff:2014kja} and exactly for a certain truncation in \cite{Sundin:2013uca}. The integrability of the remaining examples was established in \cite{Wulff:2014kja} (see also \cite{Wulff:2015mwa}) and for the non-supersymmetric solution in \cite{Wulff:2017hzy}. All these cases are in fact T-dual to $AdS_3\times S^3\times S^3\times S^1$ (or $AdS_3\times S^3\times T^4$) and the integrability also follows from this observation, however the supersymmetric cases have a Lax connection valued only in their isometry algebra, i.e. involving only 8 supercharges, which is not obvious from the duality. Beyond the construction of the Lax connection much evidence has been accumulated for the integrability of the string also at the quantum level. Most relevant to our discussion here is the formulation of an exact worldsheet S-matrix 
\cite{Arutyunov:2004vx,Staudacher:2004tk,Beisert:2005fw,Beisert:2005tm,Janik:2006dc,Beisert:2006ib,Arutyunov:2006ak,Beisert:2006ez,Arutyunov:2006yd,Borsato:2012ud,Borsato:2012ss,Borsato:2013qpa,Borsato:2013hoa,Abbott:2013mpa,Borsato:2014hja,Lloyd:2014bsa,Abbott:2014pia,Borsato:2015mma,Borsato:2016xns} which matches with perturbative worldsheet calculations up to two loops in some cases \cite{Klose:2006zd,Klose:2007wq,Klose:2007rz,Puletti:2007hq,Zarembo:2009au,Kalousios:2009ey,Rughoonauth:2012qd,Sundin:2012gc,Sundin:2013ypa,Hoare:2013pma,Bianchi:2013nra,Hoare:2013ida,Engelund:2013fja,Sundin:2014sfa,Bianchi:2014rfa,Roiban:2014cia,Sundin:2014ema,Sundin:2015uva,Bianchi:2015iza,Bianchi:2015vgw,Sundin:2016gqe}.

Integrability has also recently been ruled out for some of the backgrounds in \cite{Wulff:2017zbl}. In \cite{Wulff:2017hzy} integrability was ruled out for the $AdS_3\times S^2\times S^2\times T^3$ backgrounds not given in table \ref{tab:backgrounds} while in \cite{Wulff:2017lxh} it was ruled out for $AdS_4\times S^3\times S^3$ and its T-duals $AdS_4\times S^3\times S^2\times S^1$ and $AdS_4\times S^2\times S^2\times T^2$ as well as $AdS_3\times S^5\times H^2$, $AdS_3\times SLAG_3\times H^2$ and $AdS_3\times S^3\times S^2\times H^2$. The main result of this paper is to rule out integrability for the remaining backgrounds that have not been previously analyzed. These are indicated with an 'N' in the last column in table \ref{tab:backgrounds}.

\begin{table}[ht]
\begin{center}
\begin{tabular}{lccccc}
& & \multicolumn{2}{c}{Flux moduli} & &\\
Solution & Flux & IIA & IIB & SUSY & Integrable\\
\hline
$AdS_5\times S^5$ & RR & $\times$ & - & 32 & Y\\
$AdS_5\times SLAG_3$ & RR & $\times$ & - & - & N\\
$AdS_5\times\mathbbm{CP}^2\times S^1$ & NSNS+RR & - & $\times$ & - & Y\\
$AdS_5\times S^3\times S^2$ & RR  & $\times$ & - & - & N\\
$AdS_5\times S^2\times S^2\times S^1$ & NSNS+RR & 1 & $\times$ & - & N\\
$AdS_4\times\mathbbm{CP}^3$ & RR  & 1(-) & $\times$ & -(24) & N(Y)\\
$AdS_4\times G_{\mathbbm R}^+(2,5)$ & RR  & 1 & $\times$ & - & N\\
$AdS_4\times S^4\times S^2$ & RR  & 1 & $\times$ & - & N\\
$AdS_4\times\mathbbm{CP}^2\times S^2$ & RR  & 2 & $\times$ & - & N\\
$AdS_4\times S^2\times S^2\times S^2$ & RR  & 3 & $\times$ & - & N\\
$AdS_3\times\mathbbm{CP}^2\times S^2(H^2)\times S^1$ & NSNS+RR & 1 & $\times$ & - & N\\
$AdS_3\times\mathbbm{CP}^2\times T^3$ & NSNS+RR & - & $\times$ & - & N\\
$AdS_3\times S^3\times S^3\times S^1$ & NSNS+RR & 2 & 2 & 16 & Y\\
$AdS_3\times S^3\times S^2\times T^2$ & NSNS+RR & 1 & 1 & 8 & Y\\
$AdS_3\times S^3\times T^4$ & NSNS+RR & 1 & 2 & 16 & Y\\
$AdS_3\times S^2\times S^2\times S^2(H^2)\times S^1$ & NSNS+RR & 3 & $\times$ & - & N\\
$AdS_3\times S^2\times S^2\times T^3$ & NSNS+RR & 1/1 & 1/1 & -/8 & Y\\
$AdS_3\times S^2\times T^5$ & NSNS+RR & 1 & - & 8 & Y\\
\end{tabular}
\caption{The symmetric space $AdS_n$ backgrounds for $n>2$ whose integrability was not previously ruled out. The type of flux and the number of free parameters entering the fluxes (not counting the AdS radius) are given for the type IIA and type IIB solution respectively with '-' denoting no free parameters and '$\times$' denoting that no solution exists. The number of supersymmetries preserved by the solution is also given. When there is both a type IIA and type IIB solution they are related by T-duality. $AdS_3\times S^2\times S^2\times T^3$ has a supersymmetric and a non-supersymmetric branch while $AdS_4\times\mathbbm{CP}^3$ has a subbranch which is supersymmetric. Here $SLAG_3=SU(3)/SO(3)$ and $G_{\mathbbm R}^+(2,5)=SO(5)/U(2)$. The 'N's in the last column are the main result of this paper.}
\label{tab:backgrounds}
\end{center}
\end{table}

In \cite{Wulff:2017hzy} and \cite{Wulff:2017lxh} the idea was to use the fact that an integrable 1+1 dim. QFT should display an enormously simplified S-matrix -- any scattering process should factorize into a product of 2-to-2 scattering events \cite{Arefeva:1974bk,Kulish:1975ba,Vergeles:1976ra,Schroer:1976if,Iagolnitzer:1977sw,Zamolodchikov:1978xm}.\footnote{For massive theories one can derive the factorization of scattering from the infinite number of conserved charges following from a Lax representation. This is not true in a massless theory since massless scattering in two dimensions is much more subtle. In fact there are examples of massless theories with a Lax representation where scattering does not factorize \cite{Nappi:1979ig,Fridling:1983ha,Fradkin:1984ai,Curtright:1994be}. Here we will use factorized scattering as our definition of integrability to avoid dealing with subtleties when there are massless modes present. Indeed factorized scattering generalizes straight-forwardly to the quantum theory whereas this is less clear for the conserved charges derived from the Lax connection. I thank B. Hoare for discussions regarding this point.} For consistency this requires
\begin{itemize}
	\item[(i)] no particle production, the number of incoming particles must equal the number of outgoing particles in any scattering process
	\item[(ii)] individual momenta and masses are also conserved, only internal quantum numbers can change
\end{itemize}
By expanding the action around a suitable classical string solution one obtains a 1+1 dim. QFT which must have these properties if the string is integrable. Since the QFT one obtains this way depends on the supergravity background these consistency conditions imply necessary\footnote{It is not completely obvious that scattering has to factorize when expanding around \emph{any} classical solution if the theory is integrable. However, 
thinking loosely of integrability as solvability of the model, a property which of course cannot depend on which solution one expands about, suggests that integrability cannot be broken spontaneously. Therefore infinitely many conserved charges in involution should exist regardless of which solution one expands about and consequently scattering should factorize, at least for massive modes. Certainly for the solutions considered here, a BMN or GKP string, which are expected to have natural interpretations in a would-be dual gauge theory, it is very difficult to imagine integrability without factorization of scattering but it would be desirable to have a proof.} conditions that the background has to satisfy for integrability to be present.

In \cite{Wulff:2017hzy} the requirement of absence of particle production in (the low-energy limit of) the long spinning Gubser-Klebanov-Polyakov (GKP) string was used rule out integrability for a large class of $AdS_3\times S^2\times S^2\times T^3$ backgrounds. Since there is non-zero NSNS flux it is enough to analyze scattering of bosons and we will find here that an essentially identical calculation applied to the mixed flux backgrounds in table \ref{tab:backgrounds} either rules out integrability or reduces the background to one that it T-dual to a pure RR background.

For pure RR backgrounds the situation is more complicated since the coupling to the flux is only through fermion terms. The bosonic theory by itself is a symmetric space sigma model and therefore integrable. As we have mentioned we expect the fermions to break this integrability in the absence of supersymmetry. However things are not so bad as it turns out that it is typically enough to consider the constraint that there be no 2-to-2 scattering processes with different masses for incoming and outgoing particles. Therefore one does not have to consider the more complicated 2-to-3 amplitudes involved in particle production. In \cite{Wulff:2017lxh} this idea was applied to a general $AdS_n\times M_{10-n}$ string expanded about a Berenstein-Maldacena-Nastase (BMN) solution and it was shown that integrability implies a certain condition on the RR flux. This condition by itself is not quite enough to rule out integrability for the remaining backgrounds. The reason is that this condition came only from scattering of AdS bosons and is therefore not sensitive to the geometry of $M_{10-n}$. Here we derive a condition which is sensitive to the geometry of the internal space by considering instead scattering of internal space bosons into fermions. We will show that the resulting condition, together with the one in \cite{Wulff:2017lxh}, is strong enough to rule out integrability for the remaining backgrounds in table \ref{tab:backgrounds} completing the classification of integrable symmetric space $AdS_{n>2}$ strings.

It would of course be nice to extend this to include also the $AdS_2$ case but this will have to await a classification of such symmetric space backgrounds. Several integrable examples are however known already \cite{Sorokin:2011rr,Wulff:2014kja}.

The plan of the paper is as follows. We first rule out integrability for the backgrounds with mixed flux in table \ref{tab:backgrounds} which are not T-dual to solutions with pure RR flux. This is done by requiring absence of particle production at tree-level in the low-energy limit of the GKP string following \cite{Wulff:2017hzy}. Then we derive a condition from the absence of pair-production in the BMN string where the incoming particles are massless bosons from $M_{10-n}$ and the outgoing ones are massive fermions. We show that this condition together with the corresponding condition derived in \cite{Wulff:2017lxh} is enough to rule out integrability for all remaining backgrounds in table \ref{tab:backgrounds}.

\section{Condition on NSNS flux from GKP 2-to-3 scattering}
The general analysis of \cite{Wulff:2017lxh} assumed that there was no contribution from the NSNS flux which is not the case for the mixed flux backgrounds in table \ref{tab:backgrounds}. However, for backgrounds with non-zero NSNS flux the bosonic part of the string action is not of the form of a symmetric space sigma model due to the coupling to the B-field. This means that the question of integrability is non-trivial already at the bosonic level. Therefore it will be enough to consider only the bosonic terms in the string action in this section. We will see that the requirement of integrability of the bosonic model reduces the mixed flux backgrounds in table \ref{tab:backgrounds} to be T-dual to pure RR flux backgrounds. Following \cite{Wulff:2017hzy} a quick way to arrive at the constraint on the NSNS flux is to consider scattering in the (low-energy) bosonic GKP string. Since all excitations are massless there is no constraint from two-to-two scattering but any two-to-three amplitude must vanish in an integrable theory to be compatible with factorized scattering. Computing a few such amplitudes we will arrive at the condition on $H$. The NSNS flux for the mixed flux backgrounds in table \ref{tab:backgrounds} to be analyzed takes the form
\begin{equation}
H=R^{-1}(f_1e^3\wedge e^4+f_2e^5\wedge e^6+f_3e^7\wedge e^8)\wedge dx^9\,,
\label{eq:H}
\end{equation}
with $R$ the AdS radius and the parameters suitably restricted according to
\begin{equation}
\begin{array}{ll}
AdS_5\times S^2\times S^2\times S^1 & f_1=0\\
AdS_3\times\mathbbm{CP}^2\times S^2(H^2)\times S^1 & f_1=f_2\\
AdS_3\times\mathbbm{CP}^2\times T^3 & f_1=f_2=f_3\\
AdS_3\times S^2\times S^2\times S^2(H^2)\times S^1 & \\
\end{array}
\end{equation}
We will show below that integrability requires that $f_if_j=0$ for any $i,j=1,2,3$ such that $f_i$ and $f_j$ are associated with flux on \emph{different} irreducible factors in the geometry, e.g. for the background in the second line we have $f_1f_3=f_2f_3=0$.\footnote{This agrees with the analysis in \cite{Wulff:2015mwa} where it was shown, when the B-field is on $S^2$-factors, that a sufficient condition for a Lax connection to exist is that $H_{abe}H^{cde}\propto R_{ab}{}^{cd}$, leading to the same condition on the $f_i$'s.} This either rules out integrability or reduces the backgrounds, via T-duality, to special cases of pure RR flux backgrounds as we now explain. Looking at the form of the corresponding supergravity solutions in \cite{Wulff:2017zbl} one finds the following. With say $f_3=0$ the first background becomes T-dual to $AdS_5\times S^3\times S^2$ in table \ref{tab:backgrounds}. Similarly with say $f_2=f_3=0$ one finds that the last background becomes T-dual to $AdS_3\times S^3\times S^2\times H^2$ with only RR flux, whose integrability was ruled out in \cite{Wulff:2017lxh}. Looking at the second background we find, if $f_1=f_2=0$, that the curvature of $\mathbbm{CP}^2$ vanishes and it becomes $AdS_3\times S^2\times T^5$ which is T-dual to the integrable background $AdS_3\times S^3\times T^4$ with only RR flux, while if instead $f_3=0$ it becomes T-dual to $AdS_3\times S^5\times H^2$ with only RR flux whose integrability was again ruled out in \cite{Wulff:2017lxh}. Finally, for the third background integrability is ruled out since the $f$'s cannot all vanish.

It remains only to prove our assertion that the NSNS flux in (\ref{eq:H}) can only have legs on at most one of the curved factors in the geometry for integrability. The calculations are essentially identical to those of \cite{Wulff:2017hzy}. The bosonic terms in the string Lagrangian are
\begin{equation}
\mathcal L_b=
-\tfrac{T}{2}e_i{}^ae_j{}^b(\gamma^{ij}\eta_{ab}-\varepsilon^{ij}B_{ab})\,,
\label{eq:Lb}
\end{equation}
where $\gamma^{ij}=\sqrt{-h}h^{ij}$ with $h_{ij}$ the worldsheet metric with signature $(-,+)$ and $\varepsilon^{01}=1$. The supergravity background is encoded in the vielbeins $e^a$ ($a=0,\ldots,9$) and the B-field, which for the cases considered here is
\begin{equation}
B=R^{-1}x^9(f_1e^3\wedge e^4+f_2e^5\wedge e^6+f_3e^7\wedge e^8)\,.
\label{eq:B}
\end{equation}
When expanding this action around the long spinning (GKP) string solution one finds that the transverse AdS modes are massive. Our analysis simplifies considerably if we go to the low-energy limit where we integrate out all the massive modes. Since we are ignoring fermions this amounts to simply dropping the transverse AdS modes. In fact, in the low-energy limit, we can fix conformal gauge and use the Virasoro constraints to remove the longitudinal AdS modes as well.\footnote{This would not be true away from the low-energy limit. In that case a better approach is to consider the so-called null cusp solution related to the GKP string by analytic continuation \cite{Kruczenski:2007cy}. There one can fix the so-called AdS light-cone gauge which allows solving the Virasoro constraints. Computing scattering in this theory one finds the same results modulo terms of higher powers in the momenta which are sub-leading at low energies.} The modes associated with the compact space $M_{10-n}$ are all massless so the low-energy effective action for the GKP string (forgetting about fermions) is simply a sigma model on $M_{10-n}$ plus the B-field in (\ref{eq:B}).

For a symmetric space the vielbeins have an expansion $e^a=dx^a+\mathcal O(x^3)$ with no quadratic term, see the discussion in the next section. The only interaction terms that can contribute to 2-to-3 scattering processes at tree-level are the quartic terms from the metric and the cubic and quintic terms from the B-field. Let us consider a scattering process $x_5x_6\rightarrow x_7x_7x_9$. Since $x_9$ is the coordinate of the $S^1$ its only couplings are through the B-field in (\ref{eq:B}) and since $x_7$ belongs to a different factor in the geometry than $x_5,x_6$ it cannot couple to these through the $\mathcal O(x^3)$ terms in the vielbeins. This means that the quartic and quintic interaction terms cannot contribute and the only Feynman diagram contribution takes the form
$$
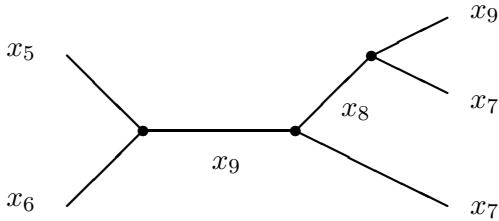
\setlength{\unitlength}{1cm}
\begin{picture}(17,5)(-1.5,0)
\thicklines
\put(3.2,3){$x_5$}
\put(3.2,1){$x_6$}
\put(4,3){\line(1,-1){1}}
\put(4,1){\line(1,1){1}}
\put(5,2){\circle*{0.15}}
\put(5,2){\line(1,0){2}}
\put(5.9,1.5){$x_9$}
\put(7,2){\circle*{0.15}}
\put(7,2){\line(1,1){1}}
\put(7.6,2.2){$x_8$}
\put(8,3){\line(2,1){1}}
\put(8,3){\line(2,-1){1}}
\put(8,3){\circle*{0.15}}
\put(9.3,3.5){$x_9$}
\put(9.3,2.3){$x_7$}
\put(7,2){\line(2,-1){2}}
\put(9.3,0.9){$x_7$}
\end{picture}
$$
This diagram is very simple to compute and using the form of the cubic vertices coming from the B-field (\ref{eq:B}) one finds, in the kinematic region $p_{1-}=p_{2+}=p_{3-}=p_{4+}=p_{5+}=0$, that the amplitude is (rescaling $x_k\rightarrow g^{-1/2}Rx_k$ where the dimensionless coupling is $g=TR^2$)
\begin{equation}
\mathcal A(x_5x_6\rightarrow x_7x_7x_9)=\frac{i}{128g^{3/2}}f_2f_3^2p_{1+}p_{2-}\,.
\end{equation}
Since this amplitude must vanish in an integrable theory we find $f_2f_3=0$. Repeating this calculation for coordinates belonging to each pair of irreducible factors in $M_{10-n}$ gives $f_if_j=0$ for each pair as claimed.

\section{Condition on RR flux from BMN 2-to-2 scattering}
It remains to analyze the integrability of the pure RR flux backgrounds in table \ref{tab:backgrounds}. Since they have vanishing NSNS flux the bosonic string action is that of a symmetric space sigma model and therefore integrable. However the full superstring actions contains fermions and the question is what happens to the integrability once they are included. The analysis of \cite{Wulff:2017lxh} shows that typically it is lost. There a constraint on the RR fluxes was derived by expanding around the BMN string and requiring scattering of two identical AdS bosons into two fermions (of a different mass) to be trivial, a necessary condition for factorized scattering. However this condition is not strong enough for our purposes here. The reason is that it does not probe the geometry of the compact space and since the only difference between for example $AdS_5\times S^5$ and $AdS_5\times SLAG_3$ is in the geometry of the compact part, in particular the flux takes the same form, one needs to derive a constraint which probes this geometry. A simple way to do this is to consider instead scattering of two bosons from the compact part into two fermions and require that amplitude to vanish (when the masses are different). This is the approach we will take here.

The pure RR flux backgrounds in table \ref{tab:backgrounds} are of the form $AdS_n\times M_{10-n}$ with $M_{10-n}$ a Riemannian symmetric space. We will expand the string in a near-BMN expansion and a suitable form for the $AdS$ metric is
\begin{equation}
ds^2_{AdS}=R^2\left(-\left(\frac{1+\tfrac14z_I^2}{1-\tfrac14z_J^2}\right)^2dt^2+\frac{dz_I^2}{(1-\tfrac14z_J^2)^2}\right)\,,
\end{equation}
with $R$ the AdS radius and $z_I$ ($I=1,\ldots,n-1$) transverse coordinates. Since we will consider scattering involving transverse modes from $M_{10-n}$ we need to know something about its geometry. Fortunately it is not hard to find general expressions using only the fact that $M_{10-n}$ is a symmetric space. For a symmetric space $G/H$ the geometry is encoded in the Maurer-Cartan one-form
\begin{equation}
J=g^{-1}dg=\tfrac12\omega^{ab}M_{ab}+e^aP_a\,,\qquad g\in G\,,
\label{eq:MC}
\end{equation}
with $\omega^{ab}$ the spin connection and $e^a$ the vielbeins. The Lie algebra of $G$ is generated by rotations $M_{ab}$ and translations $P_a$ whose commutation relations can be derived from the Maurer-Cartan equation $dJ-J\wedge J=0$ and read (e.g. \cite{Wulff:2015mwa})
\begin{equation}
[M_{ab},M_{cd}]=\delta_{ac}M_{bd}+\ldots\,,\qquad
[M_{ab},P_c]=2\delta_{c[a}P_{b]}\,,\qquad
[P_a,P_b]=-\tfrac12R_{ab}{}^{cd}M_{cd}\,,
\label{eq:alg}
\end{equation}
with $R_{ab}{}^{cd}$ the (constant) Riemann tensor of the space. Since in general $M_{ab}$ satisfies some projection conditions, so that they generate the subgroup $H\subset G$, care should be taken that only the commutation relations consistent with these projections are kept. When $M_{ab}$ occurs only contracted with $\omega^{ab}$ or $R^{ab}$ this is automatically taken care of.

Now we make the following choice of coset representative, which singles out the $9$-direction which will be involved in our light-cone coordinates,
\begin{equation}
g=e^{x^9P_9}e^{x^mP_m}\,,
\end{equation}
where $m=n,\ldots,8$ runs over what will become the transverse directions of $M_{10-n}$. From the definition of the Maurer-Cartan form in (\ref{eq:MC}) one can then easily extract the form of the vielbeins and spin connection by expanding $g$ and using the algebra (\ref{eq:alg}). We will only need the first few orders for our calculations and they are ($a,b=m,9$)
\begin{align}
e^a=&\,
dx^a
-\tfrac16dx^kx^lx^mR_{klm}{}^a
-\tfrac12dx^9x^kx^lR_{9kl}{}^a
+\mathcal O(x^5)
\,,
\\
\omega^{ab}
=&\,
-\tfrac12dx^mx^nR_{mn}{}^{ab}
+dx^9x^mR_{m9}{}^{ab}
+\mathcal O(x^4)\,.
\end{align}
Note that our choice of coset representative has eliminated the dependence on $x^9$ making shifts of $x^9$ a manifest isometry. This is what we need to fix the BMN light-cone gauge.

Since there is no B-field along the $t,x^9$-directions (in fact there is no B-field at all) it is easy to see that there exists a BMN solution of the string equations of motion describing a point-like string moving along the geodesic given by the $x^9$-direction. Rescaling $x^9\rightarrow Rx^9$ with $R$ the AdS radius it takes the form
\begin{equation}
x^+=\tfrac12(t+x^9)=\tau\,,
\label{eq:BMN-sol}
\end{equation}
with $\tau$ the worldsheet time parameter. The idea is to expand the string action around this solution, fixing light-cone gauge, and compute the tree-level scattering amplitude\footnote{One can always decompactify the worldsheet of the closed string so that it is possible to define asymptotic states and an S-matrix in the standard way, e.g. \cite{Arutyunov:2009ga}.} for two identical bosons from $M_{10-n}$ into two fermions. As remarked earlier factorized scattering requires this amplitude to vanish unless the masses of the fermions equal those of the bosons.

The Green-Schwarz string Lagrangian takes the form 
\begin{align}
\mathcal L=&
-\tfrac{T}{2}e_i{}^ae_j{}^b(\gamma^{ij}\eta_{ab}-\varepsilon^{ij}B_{ab})
-iTe_i{}^a\,\theta\Gamma_a(\gamma^{ij}-\varepsilon^{ij}\Gamma_{11})\mathcal D_j\theta
+\mathcal O(\theta^4)\,,
\label{eq:L}
\end{align}
with $\theta\Gamma^a\mathcal D\theta=\theta^\alpha\mathcal C_{\alpha\beta}(\Gamma^a)^\beta{}_\gamma(\mathcal D\theta)^\gamma$ where the Killing spinor derivative operator is
\begin{equation}
\mathcal D=d-\tfrac14\omega^{ab}\Gamma_{ab}+\tfrac18e^a(H_{abc}\Gamma^{bc}\Gamma_{11}+\mathcal S\Gamma_a)\,.
\label{eq:D}
\end{equation}
The quadratic terms in $\theta$ were determined in \cite{Cvetic:1999zs} but we will use the conventions of \cite{Wulff:2013kga} where also the $\theta^4$-terms were determined though we won't need them here. The bosonic fields appearing here are the pull-backs to the worldsheet of type II supergravity fields -- the vielbeins $e^a$ ($a=0,\ldots,9$) and spin connection $\omega^{ab}$, the NSNS two-form $B_{ab}$ and its field strength $H=dB$ and the RR field strengths encoded in the bispinor $\mathcal S$. We have written the expressions appropriate to the type IIA string with $\theta$ a 32-component Majorana spinor and
\begin{equation}
\mathcal S=e^\phi(F^{(0)}+\tfrac12F^{(2)}_{ab}\Gamma^{ab}\Gamma_{11}+\tfrac{1}{4!}F^{(4)}_{abcd}\Gamma^{abcd})\,,
\label{eq:SIIA}
\end{equation}
in terms of the dilaton $\phi$ and RR field strengths. The Lagrangian for the type IIB string is obtained by the replacements $\Gamma^a\rightarrow\gamma^a$, $\Gamma_{11}\rightarrow\sigma^3$ and
\begin{equation}
\mathcal S=-e^\phi(F^{(1)}_ai\sigma^2\gamma^a+\tfrac{1}{3!}F^{(3)}_{abc}\sigma^1\gamma^{abc}+\tfrac{1}{2\cdot5!}F^{(5)}_{abcde}i\sigma^2\gamma^{abcde})\,,
\label{eq:SIIB}
\end{equation}
where $\theta$ now consists of two 16-component Majorana-Weyl spinors of the same chirality and $\gamma^a$ are 16-component gamma matrices, while the Pauli matrices mix the two spinors.

We now expand this action around the BMN solution (\ref{eq:BMN-sol}) fixing the so-called uniform light-cone gauge
\begin{equation}
x^+=\tau\,,\qquad\frac{\partial\mathcal L}{\partial\dot x^-}=-2g\,,\qquad\frac{\partial\mathcal L}{\partial x'^-}=0\,,
\end{equation}
where we have introduced the dimensionless coupling $g=TR^2$, and kappa symmetry gauge
\begin{equation}
\Gamma^+\theta=0\qquad\Leftrightarrow\qquad \theta=P_+\theta\,,\qquad P_\pm=\tfrac12(1\pm\Gamma^{09})\,.
\end{equation}
One then solves the constraints on the momentum conjugate to $x^-$ for the worldsheet metric $\gamma^{ij}$ and the Virasoro constraints for $x^-$ itself. To simplify things we will only keep the terms in the Lagrangian which are relevant for computing the amplitude for $xx\rightarrow\theta\theta$ scattering at tree-level. Since there are no couplings of the form $zxx$ we can simply set the transverse AdS modes $z_I$ to zero. From the constraints one finds that the worldsheet metric must take the form $\gamma^{ij}=\eta^{ij}+\hat\gamma^{ij}$ with
\begin{equation}
\hat\gamma^{00}=\hat\gamma^{11}=-\tfrac12x^kx^lR_{9kl9}+\mathcal O(x^3)
\,,\qquad
\hat\gamma^{01}
=\mathcal O(x^3)\,.
\end{equation}
Rescaling the fields as $x_m\rightarrow g^{-1/2}Rx_m$, $\theta\rightarrow \frac12g^{-1/2}R^{1/2}\theta$ to make them dimensionless with canonical kinetic terms the Lagrangian becomes (recall that the B-field vanishes)
\begin{align}
\label{eq:L1}
\mathcal L=&
\tfrac12\partial_+x^m\partial_-x^n\delta_{mn}
-\tfrac12x^kx^lc_{kl}
-\tfrac{i}{2}\theta_+\Gamma^-\partial_+\theta_+
-\tfrac{i}{2}\theta_-\Gamma^-\partial_-\theta_-
-\theta_+\Gamma^{01}M\theta_-
\\
&{}
-\tfrac13c_{klm}(\partial_+x^k+\partial_-x^k)x^lx^m
-\tfrac{i}{2\sqrt g}\partial_+x^m\,\theta_+\Gamma^-N\Gamma_m\theta_-
-\tfrac{i}{2\sqrt g}\partial_-x^m\,\theta_-\Gamma^-N\Gamma_m\theta_+
\nonumber\\
&{}
+\tfrac{i}{8\sqrt g}c_{klm}x^m(\theta_+\Gamma^{-kl}\theta_++\theta_-\Gamma^{-kl}\theta_-)
-\tfrac{i}{8g}c_{kl}x^kx^l(\theta\Gamma^-\partial_-\theta_++\theta\Gamma^-\partial_+\theta_-)
\nonumber\\
&{}
+\tfrac{1}{4g}\partial_+x^m\partial_-x^n\,\theta_-\Gamma^-\Gamma_mM\Gamma^1\Gamma_n\theta_+
-\tfrac{i}{16g}[2c_{kn}\partial_-x^mx^k+c_{klmn}\partial_+x^kx^l]\,\theta_+\Gamma^{-mn}\theta_+
\nonumber\\
&{}
-\tfrac{i}{16g}[2c_{kn}\partial_+x^mx^k+c_{klmn}\partial_-x^kx^l]\,\theta_-\Gamma^{-mn}\theta_-
-\tfrac{i}{4g}c_{klm}x^kx^m(\theta_+\Gamma^-N\Gamma^l\theta_-+\theta_-\Gamma^-N\Gamma^l\theta_+)
+\ldots
\nonumber
\end{align}
where $\theta_\pm=\frac12(1\pm\Gamma_{11})\theta$ and $\partial_\pm=\partial_0\pm\partial_1$ and the dimensionless couplings involving the curvature are
\begin{equation}
c_{kl}=R^2R_{9kl9}\,,\qquad
c_{klm}=R^2R_{klm9}\,,\qquad
c_{klmn}=R^2R_{klmn}\,.
\end{equation}
Following \cite{Wulff:2017lxh} we have introduced two matrices $M$ and $N$ defined as
\begin{equation}
P_+\mathcal SP_-=4iR^{-1}\Gamma^{01}MP_-\,,\qquad P_+\mathcal SP_+=4R^{-1}NP_+
\label{eq:MN-IIA}
\end{equation} 
and used the fact that $[\Gamma^{01},\mathcal S]=0$. They are defined so that they commute with $\Gamma^\pm$ and $\Gamma_{11}$ and as a consequence $M^T=\Gamma^1M\Gamma^1$ and $N^T=-\Gamma^1N\Gamma^1$. For the type IIB case one replaces $M\rightarrow iM$ and $N\rightarrow iN$ in the Lagrangian where now (note the difference in factors of $i$)
\begin{equation}
P_+\mathcal SP_-=4R^{-1}\gamma^{01}MP_-\,,\qquad P_+\mathcal SP_+=4iR^{-1}NP_+
\label{eq:MN-IIB}
\end{equation} 
and $M,N$ now anti-commute with $\gamma^\pm$ and $\sigma^3$ and satisfy $M^T=-\gamma^1M\gamma^1$ and $N^T=\gamma^1N\gamma^1$.

The Lagrangian (\ref{eq:L1}) is somewhat complicated leading to many contributions to a given $xx\rightarrow\theta\theta$ amplitude. But looking at the backgrounds of interest in table \ref{tab:backgrounds} we notice the following. Except for the case of $AdS_4\times\mathbbm{CP}^3$, for which the condition derived in \cite{Wulff:2017lxh} turns out to be enough, the Riemann tensor of the compact factor satisfies
\begin{equation}
R_{79}{}^{ab}=0\,.
\end{equation}
This is obvious when the compact space is a direct product since the 7 and 9 directions belong to different factors. It is also true for $G_{\mathbbm R}^+(2,5)$ and $SLAG_3$ whose curvatures can be found in the appendix of \cite{Wulff:2017zbl} provided we identify the 7,9-directions with say the 1,3 and 1,5-directions respectively as labeled there. Therefore things simplify drastically if we consider $x_7x_7\rightarrow\theta\theta$ scattering. Since the $x^3$ interaction term vanishes it is consistent to set $x_m=0$ for $m\neq7$ so that all terms involving the curvature drop out and we are left with
\begin{align}
\mathcal L=&
\tfrac12\partial_+x_7\partial_-x_7
-\tfrac{i}{2}\theta_+\Gamma^-\partial_+\theta_+
-\tfrac{i}{2}\theta_-\Gamma^-\partial_-\theta_-
-\theta_+\Gamma^{01}M\theta_-
-\tfrac{i}{2\sqrt g}\partial_+x_7\,\theta_+\Gamma^-N\Gamma^7\theta_-
\nonumber\\
&{}
-\tfrac{i}{2\sqrt g}\partial_-x_7\,\theta_-\Gamma^-N\Gamma^7\theta_+
+\tfrac{1}{4g}\partial_+x_7\partial_-x_7\,\theta_-\Gamma^-\Gamma^7M\Gamma^1\Gamma^7\theta_+
+\ldots
\label{eq:L-x7}
\end{align}
Note that the $x_7$ boson is massless.

\subsection{Tree-level $x_7x_7\rightarrow\theta\theta$ scattering}
We will now compute the amplitude for scattering of two $x_7$ bosons into two fermions. The propagator for the fermions is
\begin{equation}
\langle\theta_\pm\theta_\pm\rangle
=
\left(
\begin{array}{cc}
 k_- & -\Gamma^1M\\
-\Gamma^1M & k_+
\end{array}
\right)\frac{i\Gamma^+}{k_+k_--M^TM}
\,.
\end{equation}
The solutions to the free Dirac equation, appearing as in and out states, take the form
\begin{equation}
u^i_\pm(k)=\left(
\begin{array}{c}
\sqrt{k_-}u^i\\
-m_i^{-1}\sqrt{k_+}\Gamma^1Mu^i
\end{array}
\right)\,,\qquad
M^TMu^i=m_i^2u^i\,,\quad
u^i\Gamma^-u^j=\delta^{ij}\,,
\end{equation}
where $i,j=1,\ldots,8$ label the eight fermions with corresponding mass $m_i$. For type IIB we have $i\gamma^1$ in place of $\Gamma^1$ in these expressions.

Since the $x_7$ bosons are massless they have either $p_+=0$ or $p_-=0$ on-shell. We will work in the regime $p_{1-}=0$, $p_{2+}=0$. The contribution from the quartic interaction term in (\ref{eq:L-x7}) comes from the diagram
$$
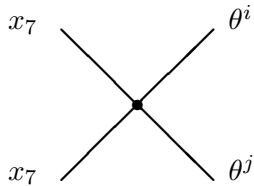
\setlength{\unitlength}{1cm}
\begin{picture}(10,3.5)(0,0)
\thicklines
\put(3.3,3){$x_7$}
\put(3.3,1){$x_7$}
\put(4,3){\line(1,-1){2}}
\put(4,1){\line(1,1){2}}
\put(5,2){\circle*{0.15}}
\put(6.2,3){$\theta^i$}
\put(6.2,1){$\theta^j$}
\end{picture}
$$
and is given by
\begin{equation}
\mathcal A_4^{ij}=
\tfrac{i}{4g}p_{1+}p_{2-}
\big(
-m_i^{-1}\sqrt{p_{3+}p_{4-}}u^iM^T\Gamma^-\Gamma^7M^T\Gamma^7u^j
+m_j^{-1}\sqrt{p_{3-}p_{4+}}u^jM^T\Gamma^-\Gamma^7M^T\Gamma^7u^i
\big)
%
\end{equation}
while the contribution from the cubic interactions is given by the sum of a t-channel and u-channel diagram
$$
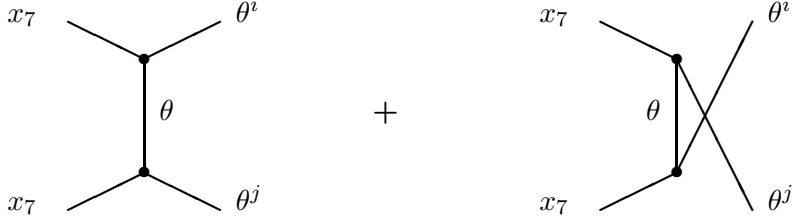
\setlength{\unitlength}{1cm}
\begin{picture}(17,5)(-1.5,0)
\thicklines
\put(1.2,3.5){$x_7$}
\put(1.2,1){$x_7$}
\put(2,3.5){\line(2,-1){1}}
\put(2,1){\line(2,1){1}}
\put(3,1.5){\line(0,1){1.5}}
\put(3,1.5){\circle*{0.15}}
\put(3,3){\circle*{0.15}}
\put(3,3){\line(2,1){1}}
\put(3,1.5){\line(2,-1){1}}
\put(4.2,3.5){$\theta^i$}
\put(4.2,1){$\theta^j$}
\put(3.2,2.2){$\theta$}
\put(6,2.2){\bf+}
\put(8.2,3.5){$x_7$}
\put(8.2,1){$x_7$}
\put(9,3.5){\line(2,-1){1}}
\put(9,1){\line(2,1){1}}
\put(10,1.5){\line(0,1){1.5}}
\put(10,1.5){\circle*{0.15}}
\put(10,3){\circle*{0.15}}
\put(10,3){\line(1,-2){1}}
\put(10,1.5){\line(1,2){1}}
\put(11.2,3.5){$\theta^i$}
\put(11.2,1){$\theta^j$}
\put(9.6,2.2){$\theta$}
\end{picture}
$$
and takes the form
\begin{equation}
\mathcal A_3^{ij}=
\tfrac{i}{4g}p_{1+}p_{2-}\big(u^i\Gamma^-\mathcal M'(p_3,p_4)u^j-u^j\Gamma^-\mathcal M'(p_4,p_3)u^i\big)
\end{equation}
where
\begin{align}
\mathcal M'
=\,&
\Gamma^1N^T\Gamma^{71}\frac{(p_1-p_3)_+\sqrt{p_{3-}p_{4-}}}{-(p_1-p_3)^2-M^TM}N\Gamma^7
+\Gamma^1M\Gamma^{71}NM\frac{m_i^{-1}\sqrt{p_{3+}p_{4-}}}{-(p_1-p_3)^2-M^TM}N\Gamma^7
\nonumber\\
&{}
-N\Gamma^{71}M\frac{m_j^{-1}\sqrt{p_{3-}p_{4+}}}{-(p_1-p_3)^2-M^TM}\Gamma^{71}NM
+M^T\Gamma^7N\Gamma^1\frac{(m_im_j)^{-1}(p_1-p_3)_-\sqrt{p_{3+}p_{4+}}}{-(p_1-p_3)^2-M^TM}\Gamma^{71}NM\,.
\end{align}
Note that the above expressions take the same form, modulo $\Gamma\rightarrow\gamma$, in the type IIA and IIB case as we have written them.

Evaluating the total amplitude on-shell and setting the fermion masses to be equal $m_i=m_j=m$ for simplicity we find
\begin{equation}
\mathcal A^{ij}=
%
%
%
%
\tfrac{i}{4mg}s
\big(
\sqrt{s-4m^2}\,u^{(i}\Gamma^-\mathcal M_S(s)u^{j)}
+\sqrt{s}\,u^{[i}\Gamma^-\mathcal M_A(s)u^{j]}
\big)
%
%
%
%
\end{equation}
where $s=-(p_1+p_2)^2=p_{1+}p_{2-}$ is the Mandelstam variable i.e. the center-of-mass energy squared,
\begin{align}
\mathcal M_S(s)
=\,&
\Gamma^1M\Gamma^{71}M^T\Gamma^7
-m^2\Gamma^1N^T\Gamma^{71}\frac{m^2-M^TM}{(m^2-M^TM)^2+sM^TM}N\Gamma^7
\nonumber\\
&{}
+\Gamma^1M\Gamma^{71}NM\frac{m^2-M^TM-s}{(m^2-M^TM)^2+sM^TM}N\Gamma^7
\nonumber\\
&{}
+N\Gamma^{71}M\frac{m^2-M^TM}{(m^2-M^TM)^2+sM^TM}\Gamma^{71}NM
\nonumber\\
&{}
+M^T\Gamma^7N\Gamma^1\frac{m^2-M^TM}{(m^2-M^TM)^2+sM^TM}\Gamma^{71}NM
\end{align}
and
\begin{align}
\mathcal M_A(s)
=\,&
\Gamma^1M\Gamma^{71}M^T\Gamma^7
-m^2\Gamma^1N^T\Gamma^{71}\frac{m^2+M^TM}{(m^2-M^TM)^2+sM^TM}N\Gamma^7
\nonumber\\
&{}
+\Gamma^1M\Gamma^{71}NM\frac{3m^2-M^TM-s}{(m^2-M^TM)^2+sM^TM}N\Gamma^7
\nonumber\\
&{}
+N\Gamma^{71}M\frac{m^2+M^TM}{(m^2-M^TM)^2+sM^TM}\Gamma^{71}NM
\nonumber\\
&{}
+M^T\Gamma^7N\Gamma^1\frac{m^2+M^TM}{(m^2-M^TM)^2+sM^TM}\Gamma^{71}NM\,.
\end{align}
Taking the limit of large center-of-mass energy, $s\rightarrow\infty$, the vanishing of the amplitude, for $m\neq0$, becomes the condition
\begin{equation}
u^i\Gamma^-M^T\Gamma^7M^T\Gamma^7u^j
-u^i\Gamma^-M^T\Gamma^7NM\frac{1}{M^TM}N\Gamma^7u^j
=0\,.
\label{eq:int-cond}
%
%
%
\end{equation}
This condition takes a similar form to the one derived in \cite{Wulff:2017lxh}. As noted there the LHS can be written without the $u$'s as $M^T\Gamma^7M^T\Gamma^7-M^T\Gamma^7NM\frac{1}{M^TM}N\Gamma^7$ but in that case one must remember to remove the projection onto fermions of mass zero. This condition was derived for equal fermion masses but it is not hard to see that the condition remains the same in the general case of unequal masses\footnote{The general solution to the energy-momentum conservation equations is, for $p_{1-}=p_{2+}=0$, $p_{3+}-p_{4+}=\sqrt{[(m_j^2-m_i^2)/p_{2-}-p_{1+}]^2-4m_i^2p_{1+}/p_{2-}}$.}. We will now see that this condition for factorization of scattering, together with the one derived in \cite{Wulff:2017lxh} from scattering of AdS bosons into fermions, is enough to rule out integrability for the remaining backgrounds in table \ref{tab:backgrounds}.

\subsection{$AdS_5\times SLAG_3$ and $AdS_5\times S^3\times S^2$}
Both these type IIB backgrounds have RR bispinor \cite{Wulff:2017zbl}\footnote{The AdS radius is related to the curvature quoted there, i.e. the proportionality factor between Ricci tensor and the metric, as $R^{-2}=\mathrm{curv}_{AdS_n}/(1-n)$.}
\begin{equation}
\mathcal S=2iR^{-1}\sigma^2(\gamma^{01234}-\gamma^{56789})\,.
\end{equation}
From the definition in (\ref{eq:MN-IIB}) we find
\begin{equation}
M=\tfrac{i}{2}\sigma^2(\gamma^{234}-\gamma^{15678})\,,\qquad N=0\,.
\end{equation}
Since $M^TMu^i=\frac12(1-\gamma^{12345678})u^i=\frac12(1-\gamma^{0123456789})u^i=u^i$ we conclude that all fermions have mass 1. Since the AdS bosons also have mass 1 their scattering into two fermions is allowed and the condition derived in \cite{Wulff:2017lxh} does not apply. However the condition derived in this section for vanishing amplitude for $x_7x_7\rightarrow\theta\theta$ applies since $x_7$ is massless. The LHS of the condition (\ref{eq:int-cond}) becomes simply
\begin{equation}
u^i\Gamma^-M^T\Gamma^7M^T\Gamma^7u^j=\delta^{ij}\,,
\end{equation}
which clearly doesn't vanish. This rules out integrability for these two backgrounds.

\subsection{$AdS_4\times M_6$}
These, in general massive, type IIA backgrounds all have a RR bispinor of the form \cite{Wulff:2017zbl}
\begin{equation}
\mathcal S=f_4+f_0(f_1\Gamma^{45}+f_2\Gamma^{67}+f_3\Gamma^{89})\Gamma_{11}-f_0\Gamma^{0123}+f_{12}\Gamma^{4567}+f_{13}\Gamma^{4589}+f_{23}\Gamma^{6789}\,,
\end{equation}
with the parameters related by the equations ($f_0\neq0$)
\begin{align}
&\qquad f_0^2(1-3[f_1^2+f_2^2+f_3^2])=5f_4^2+f_{12}^2+f_{13}^2+f_{23}^2\,,\nonumber\\
f_{12}=\,&f_4(2f_1f_2+f_3[1+f_1^2+f_2^2-f_3^2])/(1-f_1^2-f_2^2-f_3^2-2f_1f_2f_3)\,,\label{eq:f-rel}\\
f_{13}=\,&f_4(2f_1f_3+f_2[1+f_1^2-f_2^2+f_3^2])/(1-f_1^2-f_2^2-f_3^2-2f_1f_2f_3)\,,\nonumber\\
f_{23}=\,&f_4(2f_2f_3+f_1[1-f_1^2+f_2^2+f_3^2])/(1-f_1^2-f_2^2-f_3^2-2f_1f_2f_3)\,.\nonumber
\end{align}
Note that from the first equation it follows that $f_1^2+f_2^2+f_3^2\leq\frac13$. The AdS radius is given by the expression
\begin{equation}
R^{-2}=\frac{1}{12}\big(f_0^2(1+f_1^2+f_2^2+f_3^2)+f_4^2+f_{12}^2+f_{13}^2+f_{23}^2\big)\,.
\label{eq:Rminus2}
\end{equation}
The possible choices of $M_6$ with the corresponding additional relations between the parameters are
\begin{equation}
\begin{array}{ll}
\mathbbm{CP}^3 & f_1=f_2=f_3\\
G_{\mathbbm R}^+(2,5) & f_1=f_2=f_3\\
S^4\times S^2 & f_1=f_2=0\\
\mathbbm{CP}^2\times S^2 & f_1=f_2\\
S^2\times S^2\times S^2 & \\
\end{array}
\label{eq:M6}
\end{equation}
From the definition in (\ref{eq:MN-IIA}) we find
\begin{align}
M=\,&\frac{iR}{4}\Gamma^{23}\big(f_0-f_0f_3\Gamma^{4567}-f_{13}\Gamma^{123458}-f_{23}\Gamma^{123678}\big)\,,\nonumber\\
N=\,&\frac{R}{4}\big(f_4+f_0(f_1-f_2\Gamma^{4567})\Gamma^{123678}+f_{12}\Gamma^{4567}\big)\,,
\end{align}
which implies
\begin{equation}
M^TM=\frac{R^2}{16}\big(f_0^2(1-f_3\Gamma^{4567})^2+(f_{13}-f_{23}\Gamma^{4567})^2\big)\,.
\end{equation}
Decomposing the 8 physical fermions as
\begin{equation}
u^{\pm\pm\pm}=\tfrac12(1\pm\Gamma^{2345})\tfrac12(1\pm\Gamma^{2367})\tfrac12(1\pm\Gamma^{1346})u
\end{equation}
their masses are given by
\begin{align}
&m_{++\pm}^2=m_{--\pm}^2=m_1^2=\frac{R^2}{16}\big(f_0^2(1+f_3)^2+(f_{13}+f_{23})^2\big)\,,\nonumber\\
&m_{+-\pm}^2=m_{-+\pm}^2=m_2^2=\frac{R^2}{16}\big(f_0^2(1-f_3)^2+(f_{13}-f_{23})^2\big)\,.
\label{eq:masses}
\end{align}

We start by analyzing the condition derived in \cite{Wulff:2017lxh} from scattering of two mass 1 AdS bosons into fermions. The condition takes the form
\begin{equation}
m^2\delta^{ij}+u^i\Gamma^-NM\frac{1}{M^TM}NMu^j=0\,,
\label{eq:int-cond-AdS}
\end{equation}
provided that $m_i=m_j=m\neq1$. Computing the LHS with $i=++-$ and $j=+++$ we find
\begin{align}
&\frac{2f_0}{m_1^2}\left(\frac{R}{4}\right)^4\big[f_0^2(1+f_3)(f_1+f_2)-(f_4-f_{12})(f_{13}+f_{23})\big]
\nonumber\\
&\qquad\times
\big[(1+f_3)(f_4-f_{12})+(f_1+f_2)(f_{13}+f_{23})\big]u^{++-}\Gamma^-\Gamma^{123678}u^{+++}\,.
\end{align}
The only way this can vanish is if either factor in square brackets vanishes giving either
\begin{equation}
(i)\quad f_1+f_2=(f_{13}+f_{23})(f_4-f_{12})/f_0^2(1+f_3)\,,\qquad (ii)\quad f_4-f_{12}=-(f_1+f_2)(f_{13}+f_{23})/(1+f_3)\,.
\label{eq:cond-i-ii}
\end{equation}
Let us assume for the moment that $m_1\neq1$. Then we find, by taking $i=j=+++$ in the condition (\ref{eq:int-cond-AdS}), that the LHS becomes
\begin{align}
\frac{1}{m_1^2}\left(\frac{R}{4}\right)^4
\Big(
&
\left[f_0^2(1+f_3)^2+(f_{13}+f_{23})^2\right]^2
+\left[f_0^2(1+f_3)^2-(f_{13}+f_{23})^2\right]\left[(f_4-f_{12})^2-f_0^2(f_1+f_2)^2\right]
\nonumber\\
&{}
+4f_0^2(1+f_3)(f_1+f_2)(f_4-f_{12})(f_{13}+f_{23})
\Big)\,.
%
\end{align}
Combined with the previous two equations vanishing of this expression implies either
\begin{align}
(i)&\quad
\left[f_0^2(1+f_3)^2+(f_{13}+f_{23})^2\right]^2
+f_0^2(1+f_3)^2(f_4-f_{12})^2
+f_0^2(f_1+f_2)^2(f_{13}+f_{23})^2
\nonumber\\
&\qquad\qquad\qquad\qquad\qquad\qquad\qquad\qquad\qquad\qquad\qquad\qquad
+2(f_4-f_{12})^2(f_{13}+f_{23})^2
=0
\,,\nonumber\\
(ii)&\quad
\left[
f_0^4(1+f_3)^2
+2f_0^2(f_{13}+f_{23})^2
+(f_{13}+f_{23})^4/(1+f_3)^2
\right]
\left[(1+f_3)^2-(f_1+f_2)^2\right]
=0
\end{align}
Since $f_0(1+f_3)>0$ the first equation has no solution and the only solution to the second is $(1+f_3)^2-(f_1+f_2)^2=0$. Using (\ref{eq:f-rel}) and (\ref{eq:cond-i-ii}) we get $f_4[(1-f_3)^2-(f_1-f_2)^2]=0$. It is easy to show, given the previous equations and the fact that $f_1^2+f_2^2+f_3^2\leq\frac13$, that the second factor cannot vanish and we conclude that $f_4=f_{12}=f_{13}=f_{23}=0$. The first equation in (\ref{eq:f-rel}) then implies that $f_1^2+f_2^2+f_3^2=\frac13$ and adding twice this equation to $(1+f_3)^2-(f_1+f_2)^2=0$ gives $(f_1-f_2)^2+3(\frac13+f_3)^2=0$ implying that $f_2=f_1=\pm\frac13$ and $f_3=-\frac13$. Using this we find from (\ref{eq:masses}) $m_2=1$, so that there are no further constraints coming from using $u^{+-\pm}$ or $u^{-+\pm}$, while $m_1=\frac12$. Clearly there is an equivalent solution with $m_2=\frac12$ and $m_1=1$ which differs only in the sign of $f_2$ and $f_3$.

It remains only to analyze the case $m_1=m_2=1$. From $m_1=m_2$ we find using (\ref{eq:masses}) that $f_0^2f_3+f_{13}f_{23}=0$ but then it is easy to see using (\ref{eq:Rminus2}) that $m_1=1$ has no solution.

We conclude that we get 
\begin{equation}
f_1=f_2=f_3=-\frac13\,,\qquad f_4=f_{12}=f_{13}=f_{23}=0\,,
\end{equation}
or one of them equal to $-\frac13$ and the other two $+\frac13$, but since flipping the sign of two $f_i$'s gives an equivalent solution at the level of the supergravity equations they are equivalent. Looking at (\ref{eq:M6}) we see that integrability is ruled out for $M_6=S^4\times S^2$. Furthermore $AdS_4\times\mathbbm{CP}^3$ is reduced precisely to the supersymmetric solution (see \cite{Wulff:2017zbl}) which is known to be integrable.

The condition coming from scattering of AdS bosons derived in \cite{Wulff:2017lxh} is not sufficient for ruling out integrability of the remaining backgrounds completely. The reason is of course that it does not resolve the geometry of $M_6$ and therefore does not distinguish between for example $\mathbbm{CP}^3$ and $G_{\mathbbm R}^+(2,5)$. However looking at the condition coming from scattering of massless $x_7$ bosons into fermions (which as we have seen are all massive) in (\ref{eq:int-cond}) we find using the constraints derived above and taking $i,j=+++$ that the LHS becomes simply $f_0^2R^2/18=1/2$. This rules out integrability for the remaining cases.

\bibliographystyle{nb}
\bibliography{biblio}{}

\end{document}